\documentclass[english,aps,pre,reprint,showpacs,amsfonts,amsmath,amssymb,nofootinbib]{revtex4-1}
\usepackage[english]{babel}
\usepackage{graphicx}
\usepackage[dvips]{color}
\usepackage{mathrsfs}
\usepackage{enumerate}
\usepackage{color}

\begin{document}

\def\arXiv#1#2#3#4{{#1}, {\it #4} (#2).}
\def\Book#1#2#3#4#5{{#1}, {\it #3} (#4, #5, #2).}
\def\Bookwd#1#2#3#4#5{{#1}, {\it #3} (#4, #5, #2)}
\def\Journal#1#2#3#4#5#6#7{{#1}, {\it #4} \textbf{#5}, #6 (#2).}

\newcommand{\dd}{\mbox{d}}
\newcommand{\EE}{\mathbb{E}}
\newcommand{\NN}{\mathbb{N}}
\newcommand{\PP}{\mathbb{P}}
\newcommand{\RR}{\mathbb{R}}
\newcommand{\TT}{\mathbb{T}}
\newcommand{\ZZ}{\mathbb{Z}}
\newcommand{\uu}{\mathbf{1}}
\newcommand{\HH}{\mathcal{H}}
\newcommand{\orN}{\omega_{\frac{r}{N}}}

\title{Quantum walk with a general coin: Exact solution and asymptotic properties}%
\author{Miquel Montero}
\email[E-mail: ]{miquel.montero@ub.edu}
\affiliation{Departament de F\'{\i}sica Fonamental, Universitat de Barcelona (UB), Mart\'{\i} i Franqu\`es 1, E-08028 Barcelona, Spain}
\date{\today}

\begin{abstract}

In this paper we present closed-form expressions for the wave function that governs the evolution of the discrete-time quantum walk on a line when the coin operator is arbitrary. The formulas were derived assuming that the walker can either remain put in the place or proceed in a fixed direction but never move backward, although they can be easily modified to describe the case in which the particle can travel in both directions. We use these expressions to explore the properties of magnitudes associated to the process, as the probability mass function or the probability current, even though we also consider the asymptotic behavior of the exact solution. Within this approximation, we will estimate upper and lower bounds, consider the origins of an emerging approximate symmetry, and deduce the general form of the stationary probability density of the relative location of the walker. 

\end{abstract}
\maketitle

\section{Introduction}

Quantum walks~\cite{ADZ93,NV00,ABNVW01,TM02,NK03,JK03,VA12} can be thought as the quantum-mechanical version of the classical random walk, the random process that models the trajectory of a particle that at each time step moves either leftward or rightward a fixed distance. In both variants a coin toss decides the way to go, being quantum the coin in the former case.

The origins of quantum walks are linked with the field of quantum computation, in particular with the design of quantum algorithms, because genuine quantum algorithms running on quantum computers can solve problems more efficiently than their classic analogues~\cite{PS97,FG98}. In particular, quantum walks have proven to be very well suited to develop search methods~\cite{CFG03,SKW03,AMB10,MNRS11}. However, the interest on quantum walks has exceeded this scope, and attracted the attention of many researchers from distant areas as, for example, game theory~\cite{FAJ04,BFT08,CB11}. Therefore, the richer mathematical analysis of the process we make, the wider are the possible applications.

The study of the behavior of the wave function that governs the evolution of the quantum walker when the coin-related features are arbitrary is not new~\cite{TFMK03,BCGJW04,CSL08}, and some general properties have been analyzed in great detail~\cite{CSB07,JKA12,TK12}. And while analytic formulas for the wave function in terms of integral transforms do exist~\cite{FWSN07,VNY12}, {\it a general, closed-form solution was still missing\/}. In absence of such a solution, asymptotic expressions have been also derived in the past
~\cite{NK05,BP07,GJS04,AVWW11}.

Here we revisit the problem of the quantum walker and tackle the issue of finding its complete solution by looking at it from a slightly different, not very common perspective~\cite{HM09}: we consider that the particle may either move rightward or remain still. In a previous work~\cite{MM13} we have shown how this alternative formulation (the process exhibits translational invariance) can encourage the use of computational approaches not exploited before. 

This paper is organized as follows. In Sec.~\ref{Sec_Process} we give a brief introduction to the formalism used in our study of the discrete-time quantum walk on a line. In Sec.~\ref{Sec_Exact}  we present closed-form formulas for computing the wave function in the most general situation. In Sec.~\ref{Sec_Properties} we analyze some general properties of the process with the aid of these explicit expressions.  Section~\ref{Sec_Asymptotic}  is devoted to the analysis of different asymptotic approximations for the probability mass function. Particular emphasis is made on the emergence of a seeming symmetry, and on the general expression for the stationary probability density of the relative position of the walker. Conclusions are drawn in Sec.~\ref{Sec_Conclusion}, and we have left for the appendices the most technical aspects of our derivations.

\section{The process}
\label{Sec_Process}
This section pinpoints the general framework of the unidirectional discrete-time quantum walk on a line~\cite{MM13}. 
Let $\HH_P$ be the Hilbert space of discrete particle positions in one dimension, spanned by the basis $\left\{|\Psi_n\rangle : n \in \{0\}\cup \ZZ^+\right\}$. 
Let $\HH_C$ be the Hilbert space of chirality, or ``coin" states, spanned by the orthonormal basis $\left\{|0\rangle, |1\rangle\right\}$, a qubit. A unidirectional discrete-time, discrete-space quantum walk on the Hilbert space $\HH\equiv\HH_C\otimes \HH_P$ consists of a unitary operator $\hat{U}_C$,
\begin{eqnarray}
\hat{U}_C
&\equiv& e^{i \alpha} \cos \theta |0\rangle  \langle 0| +e^{-i \varphi} \sin \theta |0\rangle  \langle 1| \nonumber \\
&+& e^{i \varphi} \sin \theta  |1\rangle  \langle 0| - e^{-i \alpha} \cos \theta  |1\rangle  \langle 1|,
\label{U_coin_gen}
\end{eqnarray}
acting on the coin state, {\it the throw of the quantum coin\/}, followed by the deterministic updating of the position depending on the qubit value: 
\begin{equation*}
\hat{B} \left(|q\rangle\otimes|\Psi_n\rangle \right)= |q\rangle\otimes|\Psi_{n+q}\rangle.
\end{equation*}
Explicitly, $\hat{B}$ is a nondecreasing shift operator defined in $\HH$, which takes the following form: 
\begin{eqnarray}
\hat{B}&\equiv& |0\rangle  \langle 0| \otimes \sum_{n=0}^{\infty} |\Psi_n\rangle\langle\Psi_n|+|1\rangle  \langle 1| \otimes \sum_{n=0}^{\infty} |\Psi_{n+1}\rangle\langle\Psi_n|,\nonumber\\
&\equiv&|0\rangle  \langle 0| \otimes \hat{I}_P+|1\rangle  \langle 1| \otimes \hat{S}_P,
\end{eqnarray} 
where $\hat{I}_P$ and $\hat{S}_P$ are the identity operator and the incremental shift operator, respectively, defined in the position space $\HH_P$. 

Based upon the above, the time-evolution operator $\hat{T}$ of the unidirectional quantum walker reads
\begin{equation}
\hat{T}\equiv \hat{B}\, \hat{U},
\label{hat_T}
\end{equation}
with
\begin{equation}
\hat{U}\equiv \hat{U}_C\otimes\hat{I}_P.
\label{hat_C}
\end{equation}
When $\hat{T}$ is applied reiteratively on the initial state of the quantum walker, $|\psi\rangle_{0}\equiv|\psi\rangle_{t=0}$, one recovers the state of the system at time $t$, $|\psi\rangle_t$, 
\begin{equation}
|\psi\rangle_t =\left[ \hat{T}\right]^{t}|\psi\rangle_{0}.
\end{equation}
In our case, as the time increases in discrete steps, we set the time units so that the variable $t$ is a nonnegative integer quantity, i.e., $t \in \{0\}\cup \ZZ^+$.

We assume that the initial position of the quantum walker is totally defined, and located at the origin:
\begin{equation}
\hat{M}_0 |\psi\rangle_{0} = |\psi\rangle_{0},
\label{well_located}
\end{equation}
where
\begin{equation}
\hat{M}_n\equiv \hat{I}_C\otimes  |\Psi_{n}\rangle\langle\Psi_n|,
\label{M_def}
\end{equation}
and $\hat{I}_C$ is the identity operator of the coin space $\HH_C$. The initial coin state is, however, a general superposition of the two possible qubit values, and thus
\begin{equation}
|\psi\rangle_{0} =\left(\cos \eta |0\rangle + e^{i \gamma}\sin \eta   |1\rangle\right) \otimes |\Psi_0\rangle.
\label{psi_zero_gen}
\end{equation}
It is well-known~\cite{TFMK03} that Eqs.~\eqref{U_coin_gen} and~\eqref{psi_zero_gen} show more mathematical degrees of freedom than those own by the physical problem. In particular, we can freely set $\alpha=0$, $\gamma=0$, and restrict $\theta\in[0,\frac{\pi}{2}]$, $\varphi \in[0,\pi]$, and $\eta \in[0,\frac{\pi}{2}]$, without losing any generality.~\footnote{Some of the intermediate expressions shown along this paper may be ill-defined when either $\theta=0$ or $\theta=\frac{\pi}{2}$. In spite of that, the final formulas we obtain are still valid in the full range.} 
Summing up, we consider in the sequel that
\begin{equation}
|\psi\rangle_{0} =\left(\cos \eta |0\rangle + \sin \eta   |1\rangle\right) \otimes |\Psi_0\rangle.
\label{psi_zero}
\end{equation}
and
\begin{eqnarray}
\hat{U}_C&=& \cos \theta |0\rangle  \langle 0| +e^{-i\varphi}\sin \theta  |0\rangle  \langle 1| \nonumber \\
&+&e^{i\varphi}\sin \theta  |1\rangle  \langle 0| -\cos \theta   |1\rangle  \langle 1|.
\label{U_coin}
\end{eqnarray}
With this choice, one can understand $\theta$ and $\varphi$ as the angular spherical coordinates of a unit-length vector in a configuration space $\RR^3$, $\vec{u}$, and represent the operator $\hat{U}_C$ in terms of the Pauli {\it operators\/} 
$\sigma_j$, $j\in\{1,2,3\}$, 
\begin{eqnarray}
\sigma_1&\equiv&  |0\rangle  \langle 1| + |1\rangle  \langle 0|,\\
\sigma_2&\equiv& -i |0\rangle  \langle 1| + i |1\rangle  \langle 0|,\\
\sigma_3&\equiv& |0\rangle  \langle 0|  -\   |1\rangle  \langle 1|,
\end{eqnarray}
through the scalar projection of the Pauli vector $\vec{\sigma}$ onto the $\vec{u}$ direction~\cite{TK12},  i.e.,
\begin{equation}
\hat{U}_C
=\sin\theta\cos\varphi\, \sigma_1+\sin\theta\sin\varphi \,\sigma_2+\cos\theta \,\sigma_3.
\end{equation}
 
Finally, note that we can always translate our results into the more conventional version of the discrete-time quantum walk, in which the $|0\rangle$ state in the qubit causes the walker to move leftward. To this end, we have to extend the position space to include the states in the negative side, $\HH^{\rm e}_P\equiv{\rm span}\{|\Psi_n\rangle: n\in \ZZ\}$, and the bidirectional state $|\psi^{\rm e}\rangle_t$ is obtained by applying $\hat{D}_t$ to $|\psi\rangle_t$,
\begin{equation*}
|\psi^{\rm e}\rangle_t=\hat{D}_t |\psi\rangle_t,
\end{equation*}
where $\hat{D}_t$ is the following time-dependent shift operator defined in $\HH^{\rm e}\equiv\HH_C\otimes\HH^{\rm e}_P$,
\begin{equation}
\hat{D}_t\equiv\hat{I}_C\otimes \sum_{n=0}^{\infty} |\Psi_{2n-t}\rangle\langle\Psi_n|.
\end{equation}
In practice, this means that for any unidirectional result, $F(n,t)$, we will have that $F(n,t)=F^{\rm e}(2n-t,t)$. 

\section{Exact solution}
\label{Sec_Exact}

Let us now introduce the wave functions $\psi_{0,1}(n,t)$, the two-dimensional projection of the walker state into the position basis:
\begin{eqnarray}
\psi_{0}(n,t)&\equiv& \langle 0|  \otimes  \langle\Psi_n| \psi\rangle_t, \label{Def_Psi_0}\\
\psi_{1}(n,t)&\equiv& \langle 1|  \otimes  \langle\Psi_n| \psi\rangle_t. \label{Def_Psi_1} 
\end{eqnarray}
The evolution operator $\hat{T}$, Eq.~\eqref{hat_T}, induces the following set of recursive equations on the wave-function components:
\begin{equation}
\psi_{0}(n,t)=\cos \theta \,\psi_{0}(n,t-1)+e^{-i\varphi}\sin \theta \,\psi_{1}(n,t-1),
\label{Rec_0}
\end{equation}
and
\begin{equation}
\psi_{1}(n,t)=e^{i\varphi}\sin \theta \,\psi_{0}(n-1,t-1)
-\cos \theta\, \psi_{1}(n-1,t-1),
\label{Rec_1}
\end{equation}
which are to be solved under the assumption that the walker is initially at $n=0$, that is, $\psi_{0}(n,0)=\cos \eta\, \delta_{n,0}$, $\psi_{1}(n,0)=\sin \eta\, \delta_{n,0}$, where $\delta_{n,t}$ is the Kronecker delta.

In Appendix~\ref{App_Exact}  we show that one can use a very similar approach to which was followed in~\cite{MM13} to answer the posed problem. Here the solution reads 
\begin{eqnarray}
\psi_{0}(n,t)&=&\frac{\cos \eta}{N}\Bigg\{\frac{1+(-1)^t}{2}+\frac{1-(-1)^t}{2}\cos\theta\nonumber \\
&+&\sum_{r=1}^{N-1}  \left[1+\frac{\cos \theta \cos  \frac{\pi r}{N}}{\cos\orN}\right]\cos\left[\phi\left(\frac{n}{t},\frac{r}{N}\right) t\right]\Bigg\}\nonumber \\
&+&\frac{e^{-i \varphi}\sin \eta\sin \theta}{N}\Bigg\{\frac{1-(-1)^t}{2}\nonumber \\
&+&\sum_{r=1}^{N-1}   \frac{1}{\cos \orN}\cos\left[\phi\left(\frac{n}{t},\frac{r}{N}\right) t+\frac{\pi r}{N}\right]\Bigg\},
\label{Sol_Psi0}
\end{eqnarray}
and
\begin{eqnarray}
\psi_{1}(n,t)&=&\frac{e^{i \varphi}\cos \eta\sin\theta}{N} \Bigg\{\frac{1-(-1)^t}{2}\nonumber \\
&+&\sum_{r=1}^{N-1}  \frac{1}{\cos \orN}\cos\left[\phi\left(\frac{n}{t},\frac{r}{N}\right) t-\frac{\pi r}{N}\right]\Bigg\}\nonumber \\
&+&\frac{\sin \eta}{N}\Bigg\{\frac{1+(-1)^t}{2}-\frac{1-(-1)^t}{2}\cos\theta\nonumber \\
&+&\sum_{r=1}^{N-1}   \left[1-\frac{\cos \theta \cos  \frac{\pi r}{N}}{\cos\orN}\right] \cos\left[\phi\left(\frac{n}{t},\frac{r}{N}\right) t\right]\Bigg\},\nonumber \\
\label{Sol_Psi1}
\end{eqnarray}
where $n\in\{0,\ldots,t\}$; the constant $N$ is any natural number greater than $t$, e.g., $N=t+1$;~\footnote{In Ref.~\cite{MM13} we discussed the virtues of the alternative choice $N=2^m$, with $m$ the smallest integer for which it holds $t<2^m$. With this setting one can resort to the fast Fourier transform to perform all the calculations, thereby greatly reducing the overall computational time, a point that may be very relevant for large values of $t$.} the angular variable $\orN$ is the only solution that the equation
\begin{equation}
\orN =\arcsin\left(\cos\theta \sin  \frac{\pi r}{N}\right)
\label{omega_r}
\end{equation}
has in the $[0,\frac{\pi}{2}]$ range; and lastly
\begin{equation}
\phi\left(\frac{n}{t},\frac{r}{N}\right)\equiv\pi \left(\frac{2n}{t}-1\right)\frac{r}{N}+ \orN.
\label{phi}
\end{equation}

\begin{figure}[htbp]
\begin{tabular}{l}
(a)\\
\includegraphics[width=1.0\columnwidth,keepaspectratio=true]{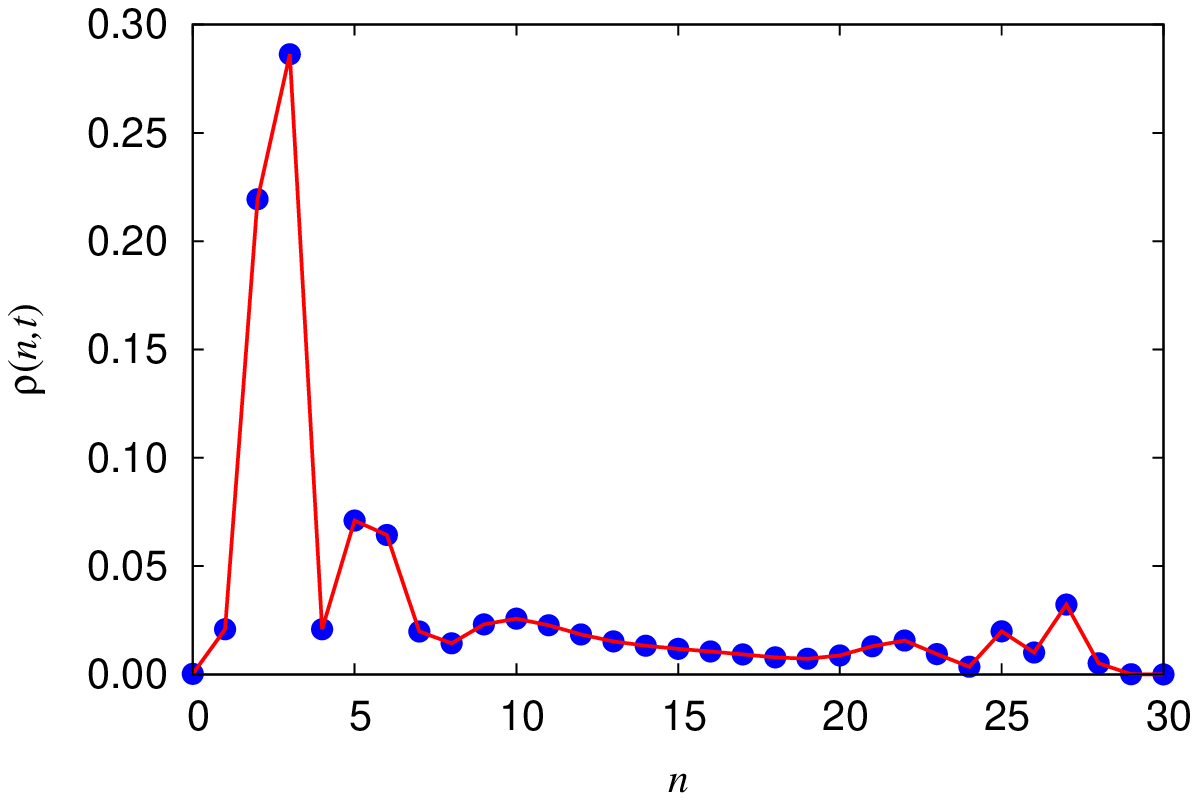}\\
(b)\\
\includegraphics[width=1.0\columnwidth,keepaspectratio=true]{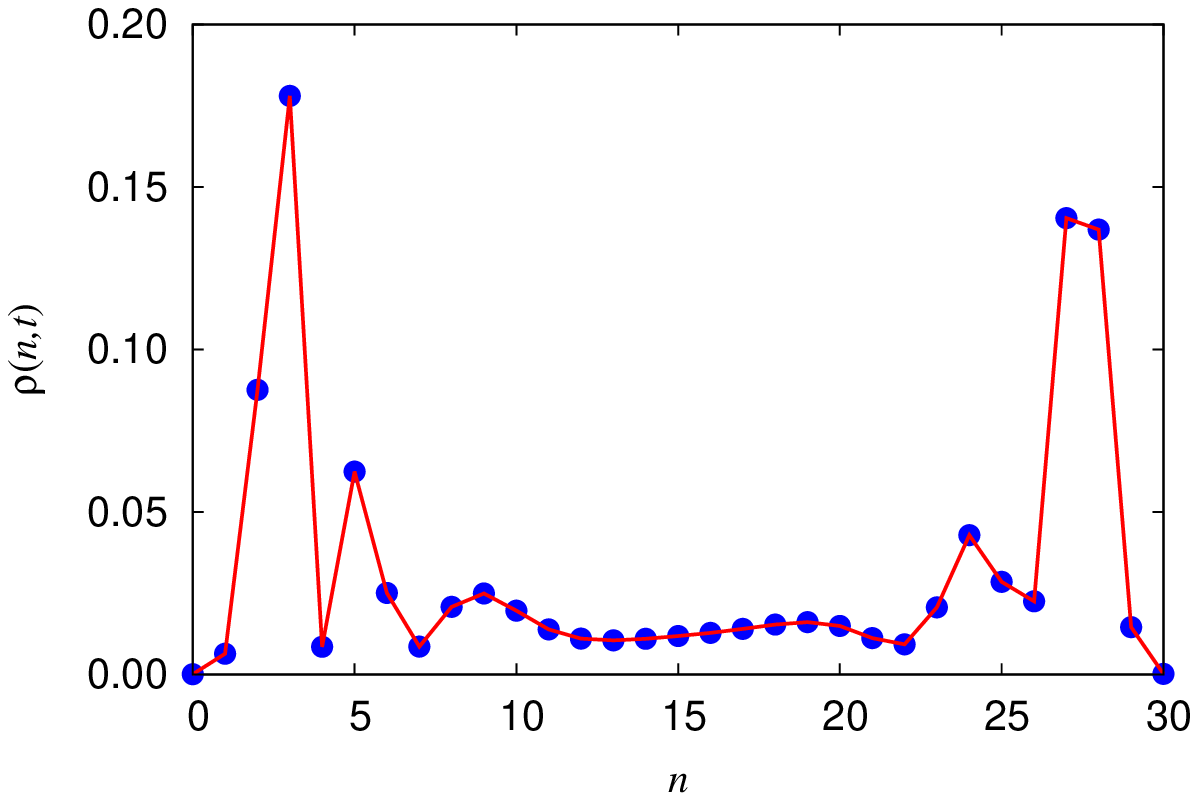}
\end{tabular}\caption{(Color online) 
Probability mass function of the process for $t=30$ time steps.  The red solid line connects the points obtained by direct application of the evolution operator on the initial state whereas the blue circles were computed employing Eq.~\eqref{rho_def_mods} when: (a) we pick the plus signs in Eq.~\eqref{U_coin_sample}; (b) we choose the minus signs in Eq.~\eqref{U_coin_sample}.} 
\label{Fig_Sample}
\end{figure}

In Fig.~\ref{Fig_Sample} we present evidence in support of the soundness of the solution shown in Eqs.~\eqref{Sol_Psi0} and~\eqref{Sol_Psi1}. For this example we have considered the outcome obtained when the coins
\begin{eqnarray}
\hat{U}_C&=& \frac{\sqrt{3}}{2} |0\rangle  \langle 0| \pm \frac{1}{2}  |0\rangle  \langle 1| \pm \frac{1}{2}  |1\rangle  \langle 0| -\frac{\sqrt{3}}{2}   |1\rangle  \langle 1|,
\label{U_coin_sample}
\end{eqnarray}
act on the following initial state:
\begin{equation}
|\psi\rangle_{0} =\left[ \frac{\sqrt{3}}{2}|0\rangle+  \frac{1}{2}|1\rangle\right] \otimes |\Psi_0\rangle.
\label{psi_zero_sample}
\end{equation}
To this end, we have computed $|\psi\rangle_t$ by systematic application of the translation operator \eqref{hat_T}, and evaluated the probability that the walker is at any given position, $\rho(n,t)$, the probability mass function (PMF) of the process, by means of
\begin{equation}
\rho(n,t)\equiv  \langle\psi|\hat{M}_n|\psi\rangle_{t} .
\label{rho_def_bracket}
\end{equation}
The results are in excellent agreement with those obtained through the numerical evaluation of  
\begin{equation}
\rho(n,t)\equiv\left|\psi_0(n,t)\right|^2+\left|\psi_1(n,t)\right|^2,
\label{rho_def_mods}
\end{equation}
for $\eta=\frac{\pi}{6}$, $\theta=\frac{\pi}{6}$, and $\varphi=0$, Fig.~\ref{Fig_Sample} (a), or with $\varphi=\pi$, Fig.~\ref{Fig_Sample} (b).

\section{General properties}
\label{Sec_Properties}
\subsection{Recursion}
The structure of Eqs.~\eqref{Sol_Psi0} and~\eqref{Sol_Psi1} shows how $\psi_{0}(n,t)$ and $\psi_{1}(n,t)$ can be seen as the superposition of the (coupled) evolution of the two initial components of the wave function, $\psi_{0}(0,0)=\cos \eta$, $\psi_{1}(0,0)=\sin \eta$. Another picture is also possible, in which formally the evolution of each component depends only on their own initial values. The price to be paid is the inclusion of the nonzero components of the wave function at time $t=1$,  
\begin{eqnarray*}
\psi_{0}(0,1)&=&\cos\eta\cos \theta+ e^{-i \varphi}\sin \eta\sin \theta,\\
\psi_{1}(1,1)&=&e^{i \varphi}\cos\eta\sin \theta- \sin \eta\cos \theta,
\end{eqnarray*}
since $\psi_0(1,1)=\psi_1(0,1)=0$. In terms of $\psi_{0}(0,0)$, $\psi_{1}(0,0)$, and the above quantities one has~\footnote{We show below how $|\psi_{1}(1,1)|^2$ can be understood as the (rightward) ``initial velocity'' of our walker. In a bidirectional scheme, $|\psi_{0}(0,1)|^2$ would play the role of the leftward ``initial velocity".}
\begin{equation}
\psi_{0}(n,t)=\psi_{0}(0,0)  \Lambda(n,t)+\psi_{0}(0,1) \Lambda(n+1,t+1),
\label{Sol_Symm_Psi0}
\end{equation}
and
\begin{equation}
\psi_{1}(n,t)=\psi_{1}(0,0) \Lambda(n,t)+\psi_{1}(1,1)\Lambda(n,t+1),
\label{Sol_Symm_Psi}
\end{equation}
where
\begin{eqnarray}
\Lambda(n,t)&\equiv&\frac{1}{N}\Bigg\{\frac{1+(-1)^t}{2}\nonumber \\
&+&\sum_{r=1}^{N-1}  \frac{1}{\cos\orN}\cos\left[\phi\left(\frac{n}{t},\frac{r}{N}\right) t-\orN\right]\Bigg\}.\nonumber \\
\label{Lambda_def}
\end{eqnarray}
The two equations that correlate the evolution of the wave functions, cf. Eqs.~\eqref{Rec_0} and~\eqref{Rec_1}, turn now into a single, two-step recursive formula that governs the whole dynamics:~\footnote{This idea of expressing the evolution of the walker in terms of a unique magnitude which follows a recursive equation is not new, e.g., in Ref.~\cite{ABNVW01} we find one of such recursion formulas which involves non-constant coefficients. Especially relevant is, in this sense, Ref.~\cite{FWSN07} where it is derived the bidirectional equivalent of Eq.~\eqref{Lambda_recursive} on the basis of the properties of the Chebyshev polynomials of the second kind.}
\begin{eqnarray}
\Lambda(n,t+2)&=&\cos \theta \left[\Lambda(n,t+1)-\Lambda(n-1,t+1)\right]\nonumber\\
&+&\Lambda(n-1,t).
\label{Lambda_recursive}
\end{eqnarray}
Expression~\eqref{Lambda_def} is recovered from Eq.~\eqref{Lambda_recursive} once one considers the following initial conditions $\Lambda(0,0)=1$, and $\Lambda(0,1)=\Lambda(1,1)=0$, together with the boundary conditions $\Lambda(-1,t)=\Lambda(t+1,t)=0$, for $t\geq 0$. In fact, these conditions lead to $\Lambda(0,t)=\Lambda(t,t)=0$, for $t\geq 1$ as well. 

\subsection{Symmetry}

Function $\Lambda(n,t)$ is very useful for the analysis of space symmetries around the middle of the wave packet~\cite{TK12,JKA12}. It is easy to check from Eq.~\eqref{Lambda_def} that one has
\begin{eqnarray*}
\Lambda(k-l,2k)&=&\Lambda(k+l,2k),\\
\Lambda(k-l,2k+1)&=&-\Lambda(k+l+1,2k+1),
\end{eqnarray*}
 for any $l\in\{0,\ldots,k\}$, $k\geq 0$. In other words, $\Lambda(n,t)$ is symmetric around the point $\frac{t}{2}$,
\begin{equation}
\Lambda\left(\left\lfloor\frac{t}{2}\right\rfloor-l,t\right)=(-1)^t\Lambda\left(\left\lceil\frac{t}{2}\right\rceil+l,t\right),
\label{Lambda_Symm}
\end{equation}
where we have introduced the floor function, $\left\lfloor\cdot\right\rfloor$, and the ceiling function,  $\left\lceil\cdot\right\rceil$. Accordingly, $\Lambda(n,t+1)$ is symmetric around the point $\frac{t+1}{2}$, while $\Lambda(n+1,t+1)$ is symmetric around $\frac{t-1}{2}$. One can be remove this disparity in the center of the symmetry with the help of the two auxiliary functions $\Lambda_{\pm}(n,t)$,
\begin{equation}
\Lambda_{\pm}(n,t)\equiv\frac{1}{2}\left[\Lambda(n+1,t+1)\pm\Lambda(n,t+1)\right],
\end{equation}
since it can be shown that
\begin{equation}
\Lambda_{+}\left(\left\lfloor\frac{t}{2}\right\rfloor-l,t\right)=(-1)^{t+1}\Lambda_{+}\left(\left\lceil\frac{t}{2}\right\rceil+l,t\right),
\label{Lambda_p_Symm}
\end{equation}
and
\begin{equation}
\Lambda_{-}\left(\left\lfloor\frac{t}{2}\right\rfloor-l,t\right)=(-1)^t\Lambda_{-}\left(\left\lceil\frac{t}{2}\right\rceil+l,t\right).
\label{Lambda_m_Symm}
\end{equation}
In terms of these three quantities $\Lambda(n,t)$ and $\Lambda_{\pm}(n,t)$ the PMF reads
\begin{eqnarray}
&&\rho(n,t)=\Lambda^2(n,t)+\Lambda_{+}^2(n,t) +\Lambda_{-}^2(n,t)\nonumber\\
&+&2\cos\theta\, \Lambda(n,t)\Lambda_{-}(n,t)\nonumber\\
&+& 2\left[\cos 2\eta \cos 2\theta+\sin 2 \eta\sin 2\theta\cos\varphi\right]\Lambda_{+} (n,t)\Lambda_{-}(n,t)\nonumber\\
&+&2\left[ \cos 2\eta \cos \theta+\sin 2 \eta\sin \theta\cos\varphi\right]\Lambda (n,t)\Lambda_{+}(n,t).\nonumber\\
\label{Prob_Symm}
\end{eqnarray}
The first three terms in the right hand side of Eq.~\eqref{Prob_Symm} are even functions around the midpoint. The same holds for the product  $\Lambda(n,t)\Lambda_{-}(n,t)$, see Eqs.~\eqref{Lambda_Symm} and~\eqref{Lambda_m_Symm}. Conversely, $\Lambda (n,t)\Lambda_{+}(n,t)$ and $\Lambda_{+} (n,t)\Lambda_{-}(n,t)$ are odd functions. Therefore, in order to get a symmetric PMF one has to demand that  
\begin{equation}
\cos 2\eta \cos 2\theta+\sin 2 \eta\sin 2\theta\cos\varphi=0,
\label{symm_step_one}
\end{equation}
as well as
\begin{eqnarray}
\cos 2\eta \cos \theta+\sin 2 \eta\sin \theta\cos\varphi=0.
\label{symm_step_two}
\end{eqnarray}
The solutions of Eqs.~\eqref{symm_step_one} and~\eqref{symm_step_two} can be classified in three different (although intersecting) families:

The first one corresponds to $\eta=\frac{\pi}{4}$ and $\theta=0$. When $\theta=0$, the coin operator does not entangle the evolution of the wave functions, and thus one has
\begin{eqnarray*}
\psi_{0}(0,1)&=&\psi_{0}(0,0)=\cos\eta,\\
\psi_{1}(1,1)&=&-\psi_{1}(0,0)=-\sin \eta,
\end{eqnarray*}
and
\begin{equation*}
 \Lambda(n,t)=(-1)^{n-1},
\end{equation*}
for $1\leq n\leq t-1$. As a consequence
\begin{eqnarray*}
\psi_{0}(n,t)&=&\cos\eta\, \delta_{n,0},\\
\psi_{1}(n,t)&=&(-1)^t \sin \eta\, \delta_{n,t},
\end{eqnarray*}
the probability is concentrated symmetrically at the endpoints, if both wave functions have the same weight at the beginning. 

The second family of solutions, $\eta=\frac{\pi}{4}$ and $\varphi=\frac{\pi}{2}$, is based on the odd symmetry that the coin operator shows under the interchange $|0\rangle \leftrightarrow |1\rangle$ when $\varphi=\frac{\pi}{2}$. 
If the initial state is invariant under the same transformation, the PMF will be symmetric~\cite{JK03}.

The third one, related to $\eta=\frac{\pi}{4}$ and $\theta=\frac{\pi}{2}$, is pathological to some extend. Condition $\theta=\frac{\pi}{2}$ causes that $\Lambda(n,t)=\delta_{2n,t}$, and therefore $\Lambda (n,t)\Lambda_{\pm}(n,t)=0$ identically. In fact,
\begin{eqnarray*}
\rho(n,t)&=&\delta_{2n,t}+\sin^2 \eta\, \delta_{2n,t-1} +\cos^2 \eta\, \delta_{2n,t+1},
\end{eqnarray*}
and the walker remains {\it confined\/} in the smallest interval containing $\frac{t}{2}$, i.e., $\rho(n,t)=0$ for $n \not\in\{\frac{t-1}{2},\frac{t}{2},\frac{t+1}{2}\}$. The symmetry is exact only for $\eta=\frac{\pi}{4}$, whereas if $\cos 2\eta=\pm 1$, the walker's position alternates between $n=\frac{t}{2}$, when $t$ is even, and $n=\frac{t\pm1}{2}$, when $t$ is odd. Eventually, the value of $\varphi$ is found to be irrelevant in this case, a fact that cannot be foreseen from Eqs.~\eqref{symm_step_one} and~\eqref{symm_step_two}.

There are no further symmetries around the central point, which is not inconsistent with the fact that the same sets of solutions for the parameters $\theta$, $\varphi$ and $\eta$ can be obtained by requiring that alternative constraints are met~\cite{NK03}. For instance, one may demand either equity in the ``initial conditions" of the wave functions,
\begin{eqnarray}
\left|\psi_0(0,0)\right|^2&=&\left|\psi_1(0,0)\right|^2,\label{initial_x}\\ 
\left|\psi_0(0,1)\right|^2&=&\left|\psi_1(1,1)\right|^2,\label{initial_v}
\end{eqnarray}
which links with the vision of the quantum walker as a ballistic process~\cite{TK12}, or the use of an {\it always-fair\/} coin,
\begin{equation}
\langle \psi |\hat{U}|\psi\rangle_{t}=0, \forall t,
\label{fair_coin_all_times}
\end{equation}
which clearly connects with the game-theory interpretation of the quantum walker~\cite{FAJ04,BFT08,CB11}.

\subsection{Probability current}

Another relevant magnitude whose description is simpler within the present formulation is the probability current $J(n,t)$~\cite{HBF03}. Like the motion of the walker itself, the probability flux is directional and enters into the next site through the $|1\rangle$ component exclusively. Therefore, it can be evaluated through the following compact expression:
\begin{equation}
J(n,t)\equiv\left|\psi_{1}(n+1,t+1)\right|^2,
\label{J_def}
\end{equation}
for $n\in\{0,\ldots,t\}$. Alternatively, 
we can also express $J(n,t)$ in terms of the {\it local\/} values of the wave function, cf. Eq.~\eqref{Rec_1},
\begin{equation}
J(n,t)\equiv \left|e^{i\varphi}\sin \theta \,\psi_{0}(n,t)-\cos \theta \,\psi_{1}(n,t)\right|^2.
\label{J_def_alternative}
\end{equation}
We can verify the validity of this identity by obtaining the continuity equation for the probability mass function $\rho(n,t)$. Consider $\Delta_t \rho(n,t)$,
\begin{eqnarray}
\Delta_t \rho(n,t)&\equiv&\rho(n,t+1)-\rho(n,t)\nonumber\\
&=&\left|\psi_0(n,t+1)\right|^2+\left|\psi_1(n,t+1)\right|^2\nonumber \\
&-&\left|\psi_0(n,t)\right|^2-\left|\psi_1(n,t)\right|^2,\nonumber
\end{eqnarray}
and see how
\begin{eqnarray*}
\left|\psi_1(n+1,t+1)\right|^2&=&\left|e^{i\varphi}\sin \theta \,\psi_{0}(n,t)-\cos \theta \,\psi_{1}(n,t)\right|^2\\
&=&\left|\psi_0(n,t)\right|^2+\left|\psi_1(n,t)\right|^2\\
&-&\left|\cos \theta \,\psi_{0}(n,t)+e^{-i\varphi}\sin \theta \,\psi_{1}(n,t)\right|^2\\
&=&\left|\psi_0(n,t)\right|^2+\left|\psi_1(n,t)\right|^2\\
&-&\left|\psi_0(n,t+1)\right|^2,
\end{eqnarray*}
where we have first used Eq.~\eqref{Rec_1} and finally Eq.~\eqref{Rec_0}. Therefore we have
\begin{eqnarray}
\Delta_t \rho(n,t)&=&\left|\psi_1(n,t+1)\right|^2-\left|\psi_1(n+1,t+1)\right|^2\nonumber \\ 
&=&J(n-1,t)-J(n,t),
\label{Continuity_def}
\end{eqnarray}
that is, the change in the probability of the walker to be found in a given location $n$ comes from the the balance between the outgoing probability, $J(n,t)$, which goes to the $n+1$ site, and the ingoing probability $J(n-1,t)$, which comes from the $n-1$ site. 

Finally, note how Eq.~\eqref{Continuity_def} gives us the possibility of deriving the equivalent of the Ehrenfest's theorem for the time evolution of the expectation value of the position, $\langle X \rangle_t$,
\begin{equation}
\langle X \rangle_t \equiv\sum_{n=0}^{t} n \rho(n,t).
\label{X_def}
\end{equation}
In our case we have
\begin{eqnarray}
\Delta_t \langle X \rangle_t&\equiv&\langle X \rangle_{t+1}-\langle X \rangle_{t}\nonumber\\
&=&\sum_{n=1}^{t+1} n\, J(n-1,t)-\sum_{n=0}^{t} n\, J(n,t)\nonumber \\
&=&\sum_{n=0}^{t} J(n,t)=\sum_{n=0}^{t+1}\left|\psi_{1}(n,t+1)\right|^2.
\label{Ehrenfest}
\end{eqnarray}

\section{Asymptotic evolution}
\label{Sec_Asymptotic}

Despite their accuracy, most of the previous formulas are too intricate to draw further conclusions from them. The usual approach to overcome this circumstance is to consider the asymptotic limit~\cite{AVWW11,MM13}, the limit in which $t\gg 1$, $n\gg 1$, but that keeps $\nu\equiv n/t$ finite. In Appendix~\ref{App_Asymptotic} we show how, under the previous assumptions, $\rho(n,t)$ can be approximated in the domain~\cite{NK03,BP07}
\begin{equation}
\frac{1}{2}\left(1-\cos\theta\right)\leq\nu\leq\frac{1}{2}\left(1+\cos\theta\right),
\label{nu_limits}
\end{equation}  
by $\bar{\rho}(n,t)$,
\begin{eqnarray}
\bar{\rho}(n,t)&\equiv&\frac{1}{2 \pi t} \frac{1}{\nu(1-\nu)}\frac{\sin\theta}{\sqrt{\cos^2\theta-(2\nu-1)^2}} \nonumber \\
&\times&\Bigg\{1-(2\nu-1) \left(\cos 2\eta +\sin 2\eta\tan\theta \cos \varphi  \right)\nonumber\\
&+&\left|R(\nu)\right| \sin \bigg[ 2 \phi_0(\nu) t+\Omega(\nu)\bigg]\Bigg\},
\label{Main_Prob_Asymp}
\end{eqnarray}
where
\begin{eqnarray}
\phi_0(\nu) &\equiv& (2\nu-1) \arcsin\left(\sqrt{\frac{\cos^2 \theta-(2\nu-1)^2}{4\nu(1-\nu)\cos^2 \theta}}\right)\nonumber\\
&+&\arcsin\left(\sqrt{\frac{\cos^2 \theta-(2\nu-1)^2}{4\nu(1-\nu)}}\right),\label{phi_zero_Main}\\ 
R(\nu)&\equiv&(2\nu-1) \Bigg[\left|\frac{2\nu-1}{\cos^2\theta}-\cos 2\eta -\sin 2\eta \tan\theta\cos \varphi \right|^2\nonumber\\
&+&\left|1-\frac{(2\nu-1)^2}{\cos^2\theta}\right|\bigg|\cos 2\eta \tan\theta-\sin 2\eta \cos \varphi\bigg|^2\Bigg]^{\frac{1}{2}},\nonumber\\
\end{eqnarray}
and
\begin{eqnarray}
\tan\Omega(\nu)&\equiv&\frac{\sqrt{1-\frac{(2\nu-1)^2}{\cos^2\theta}}\bigg[\cos 2\eta \tan\theta-\sin 2\eta \cos \varphi\bigg]}{\frac{2\nu-1}{\cos^2\theta}-\cos 2\eta -\sin 2\eta \tan\theta \cos \varphi}.\nonumber\\
\label{Omega_Main}
\end{eqnarray}

Before analyzing in detail the mathematical structure and the main properties of Eq.~\eqref{Main_Prob_Asymp}, which will be the subject of forthcoming sections, let us devote some words to the basic nature of $\bar{\rho}(n,t)$. It is clear from Eqs.~\eqref{Main_Prob_Asymp} to~\eqref{Omega_Main} that $\bar{\rho}(n,t)$ is a function that depends intrinsically on $\nu$ and $t$. We have avoided to adopt a notation that exploits this fact because it could induce to the erroneous impression that $\bar{\rho}(n,t)$ is a (perhaps approximate) probability density function with respect to $\nu$, when the true is that it  estimates $\rho(n,t)$, a probability mass function on $n$. Therefore, for the moment, consider $\nu$ as a mere shorthand for $n/t$. We will return to this issue later on.

\subsection{Bounds}
\label{Sec_Bounds}

\begin{figure}[htbp]
{
\begin{tabular}{l}
(a)\\
\includegraphics[width=1.0\columnwidth,keepaspectratio=true]{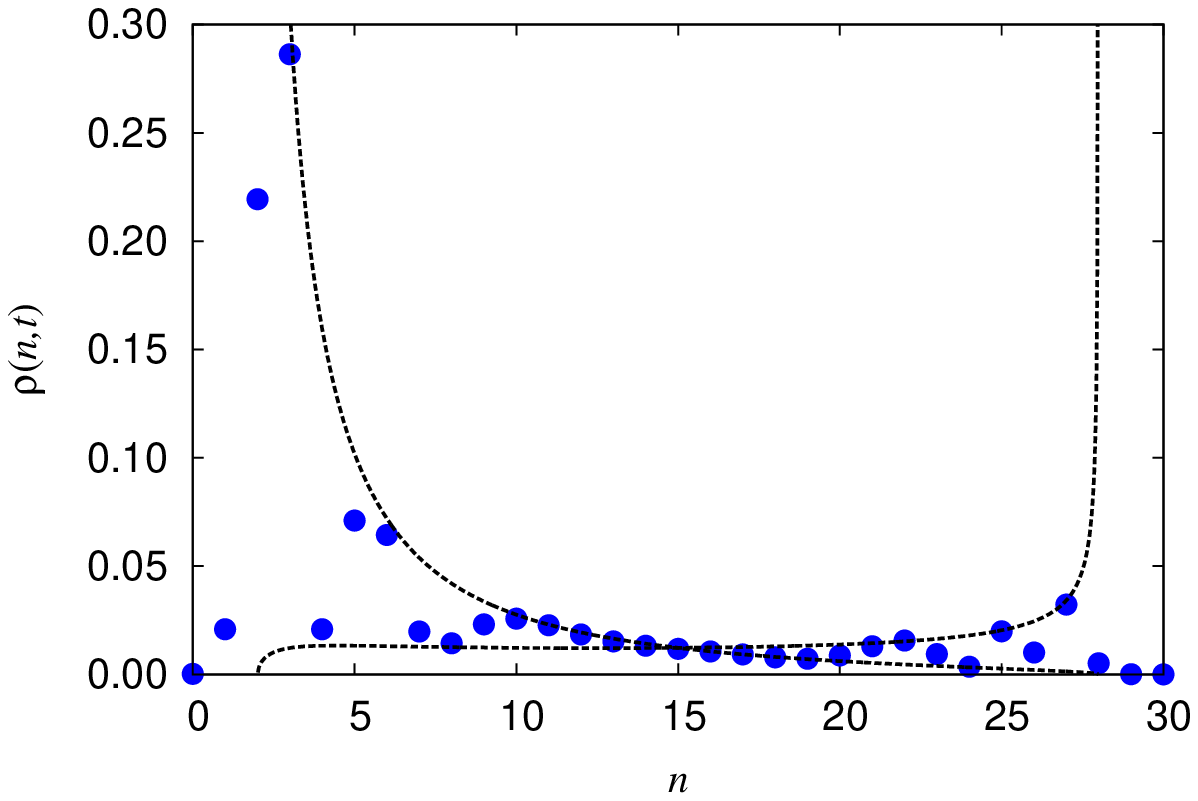}\\
(b)\\
\includegraphics[width=1.0\columnwidth,keepaspectratio=true]{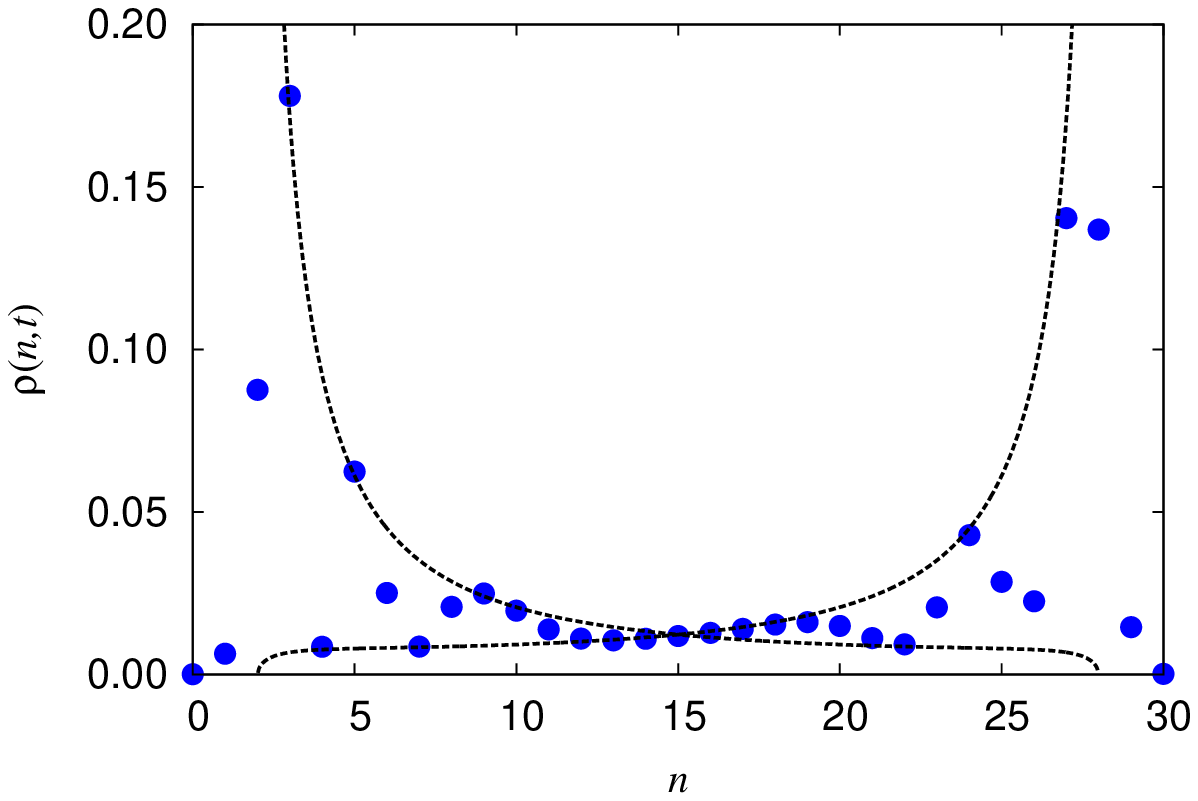}
\end{tabular}
}
\caption{(Color online) 
Probability mass function of the process for $t=30$ time steps.  The blue circles were obtained by direct evaluation of the exact solution. The upper and lower black dashed lines correspond to $\bar{\rho}_{\sup}(n,t)$ and $\bar{\rho}_{\inf}(n,t)$ respectively. 
The parameters were $\eta=\frac{\pi}{6}$, $\theta=\frac{\pi}{6}$, and $\varphi=0$ in the top panel, or $\varphi=\pi$ in the bottom panel.
} 
\label{Fig_Bounds}
\end{figure}

The first point to note is the presence of a single sinusoidal term in Eq.~\eqref{Main_Prob_Asymp}. This means that we can define $\bar{\rho}_{\sup}(n,t)$,
\begin{eqnarray}
\bar{\rho}_{\sup}(n,t)&\equiv&\frac{1}{2 \pi t} \frac{1}{\nu(1-\nu)}\frac{\sin\theta}{\sqrt{\cos^2\theta-(2\nu-1)^2}}\big[1+\left|R(\nu)\right| \nonumber \\
&-&(2\nu-1) \left(\cos 2\eta +\sin 2\eta \tan\theta\cos \varphi \right)\big], 
\label{Prob_Asymp_Max}
\end{eqnarray}
and $\bar{\rho}_{\inf}(n,t)$,
\begin{eqnarray}
\bar{\rho}_{\inf}(n,t)&\equiv&\frac{1}{2 \pi t} \frac{1}{\nu(1-\nu)}\frac{\sin\theta}{\sqrt{\cos^2\theta-(2\nu-1)^2}}\big[1-\left|R(\nu)\right|  \nonumber \\
&-&(2\nu-1) \left(\cos 2\eta +\sin 2\eta \tan\theta \cos\varphi \right)\big],
\label{Prob_Asymp_Min}
\end{eqnarray}
in such a way that $\bar{\rho}_{\inf}(n,t)\leq \bar{\rho}(n,t) \leq \bar{\rho}_{\sup}(n,t)$. This does not imply that $\bar{\rho}_{\inf}(n,t)\leq \rho(n,t) \leq \bar{\rho}_{\sup}(n,t)$, but as Fig.~\ref{Fig_Bounds} shows, these two functions are very good proxies for estimating the bounds of $\rho(n,t)$ \cite{BP07}. 

In Fig.~\ref{Fig_Bounds}  we have considered the same two particular examples we introduced in Fig.~\ref{Fig_Sample}, e.g., $\eta=\frac{\pi}{6}$, $\theta=\frac{\pi}{6}$, and either $\varphi=0$ or  $\varphi=\pi$. In both cases the upper and lower lines coincide in the middle of the plot. This is a general property since 
\begin{equation}
\lim_{n\to\frac{t}{2}}\bar{\rho}_{\inf}(n,t)=\lim_{n\to\frac{t}{2}}\bar{\rho}_{\sup}(n,t)=\frac{2}{\pi t} \tan \theta.
\end{equation} 

Another outstanding feature is that we can recover the same overall picture in Fig.~\ref{Fig_Bounds} if we replace $\bar{\rho}_{\sup}(n,t)$ and $\bar{\rho}_{\inf}(n,t)$ by two smooth functions that intersect each other at $\nu=\frac{1}{2}$. These two functions are obtained by exchanging $\left|R(\nu)\right|$ for $R(\nu)$ in Eqs.~\eqref{Prob_Asymp_Max} and~\eqref{Prob_Asymp_Min}.

Finally, for the particular case shown in the bottom panel of Fig.~\ref{Fig_Bounds}, it is clear that both $\bar{\rho}_{\sup}(n,t)$ and $\bar{\rho}_{\inf}(n,t)$ exhibit a symmetric behavior around the middle point, although this example does not belong to any of the three families of solutions of Eqs.~\eqref{symm_step_one} and~\eqref{symm_step_two} discussed above. 

\subsection{Approximate symmetry}
\label{Sec_Add_Symmetry}

The symmetry of the bounding functions $\bar{\rho}_{\sup}(n,t)$ and $\bar{\rho}_{\inf}(n,t)$ around $\nu=\frac{1}{2}$ reported in the previous section can be obtained by demanding that
\begin{equation}
\cot 2\eta =-\tan \theta \cos \varphi, 
\label{Symm_App}
\end{equation}
since then 
\begin{equation}
\bar{\rho}_{\sup \atop \inf}\mathop{(n,t)}=\frac{1}{2 \pi t} \frac{\sin\theta}{\nu(1-\nu)}\frac{1\pm\left|R(\nu)\right|}{\sqrt{\cos^2\theta-(2\nu-1)^2}},
\end{equation}
with
\begin{eqnarray}
R(\nu)&=&(2\nu-1) \Bigg[\left|\frac{2\nu-1}{\cos^2\theta}\right|^2\nonumber\\
&+&4\left|1-\frac{(2\nu-1)^2}{\cos^2\theta}\right|\cos^2 2\eta \cot^2 2\theta\Bigg]^{\frac{1}{2}}.
\end{eqnarray}
Equation~\eqref{Symm_App} coincides with Eq.~\eqref{symm_step_two}, recall that $\theta=\frac{\pi}{2}$ is a singular case, demonstrating in this way that Eqs.~\eqref{symm_step_one} and~\eqref{symm_step_two} had different level of significance. 

Several interpretations can be given to condition~\eqref{Symm_App}. Thus, in Ref.~\cite{TK12} we find a discussion about the parity properties of the quasi-momentum components of the quantum walk which ultimately would lead to Eq.~\eqref{Symm_App}. Another approach is the following: if one assumes that Eq.~\eqref{symm_step_two} holds,  $\rho(n,t)$ reduces to 
\begin{eqnarray}
\rho(n,t)&=&\Lambda(n,t)\Lambda(n+1,t+2)+\Lambda_{+}^2(n,t) +\Lambda_{-}^2(n,t)\nonumber\\
&-& 2\cos 2\eta\, \Lambda_{+} (n,t)\Lambda_{-}(n,t),
\label{Prob_Symm_App}
\end{eqnarray}
where uniquely the last term does not show even symmetry around the central point. But $\Lambda_{+} (n,t)\Lambda_{-}(n,t)$ is a second-order correction since one has
\begin{equation*}
\Lambda_{+} (n,t)\Lambda_{-}(n,t)=\frac{1}{4}\left[\Lambda^2(n+1,t+1)-\Lambda^2(n,t+1)\right].
\end{equation*}
Additional readings are: the approximate symmetry appears when one replaces Eqs.~\eqref{initial_x} and~\eqref{initial_v} with
\begin{equation}
\left|\psi_0(0,0)\right|^2+\left|\psi_0(0,1)\right|^2=\left|\psi_1(0,0)\right|^2+\left|\psi_1(1,1)\right|^2,
\end{equation}
or when one relaxes the fair-coin condition~\eqref{fair_coin_all_times}  and only asks to be fulfilled by the initial state,
\begin{equation}
\langle \psi |\hat{U}|\psi\rangle_{t=0}=0.
\label{fair_coin}
\end{equation}

Note that Eq.~\eqref{Symm_App}  has one, and only one, solution for $\eta$, given $\theta$ and $\varphi$, because the function $\cot 2\eta$ is unbounded, monotonically decreasing, and continuous for $\eta \in[0,\frac{\pi}{2}]$.  Thus, for instance, when $\theta=\frac{\pi}{4}$ and $\varphi=\pi$ one finds that $\eta=\frac{\pi}{8}$, a setup discussed in Ref.~\cite{AVWW11}.  For $\theta=\frac{\pi}{6}$ and $\varphi= \pi$, one has $\eta=\frac{\pi}{6}$, which is the case we use as an illustrative example in the bottom panels in Figs.~\ref{Fig_Sample} and~\ref{Fig_Bounds}. These plots clearly show that the symmetry is only approximate, but it becomes more and more accurate as one increases $t$: 
see Fig.~\ref{Fig_Median} (b) below. 

\subsection{Stationary density}
\label{Sec_Median}

\begin{figure}[htbp]
{
\begin{tabular}{l}
(a)\\
\includegraphics[width=1.0\columnwidth,keepaspectratio=true]{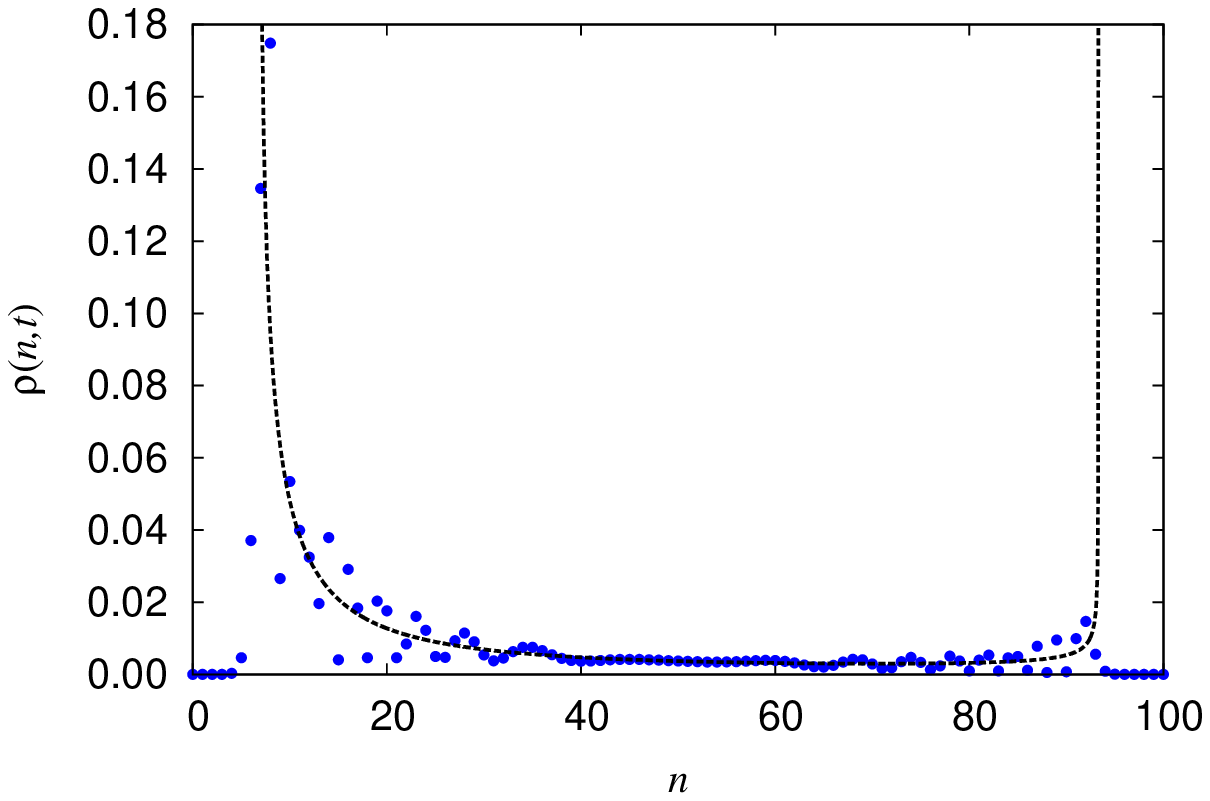}\\
(b)\\
\includegraphics[width=1.0\columnwidth,keepaspectratio=true]{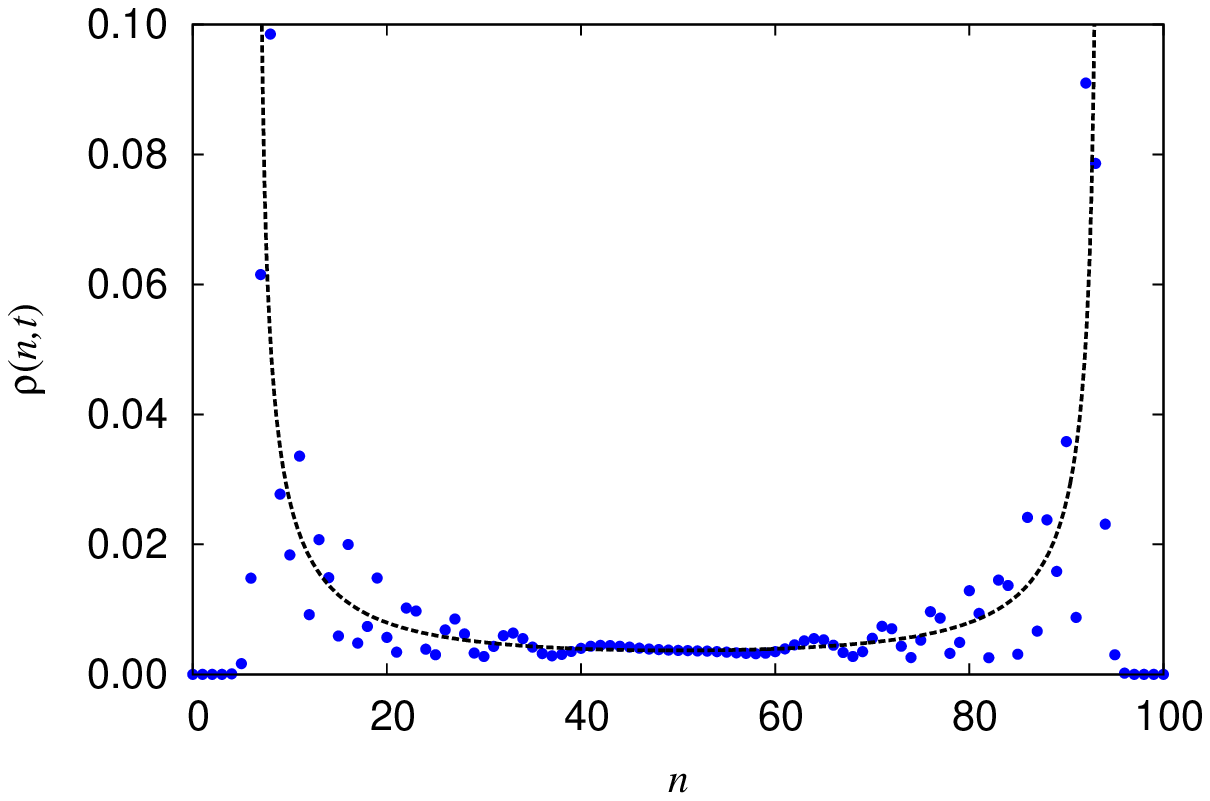}
\end{tabular}
}
\caption{(Color online) Probability mass function of the process for $t=100$ time steps.  The blue circles were obtained through the evaluation of the exact expression for $\eta=\frac{\pi}{6}$, $\theta=\frac{\pi}{6}$, and 
(a) $\varphi=0$; (b) $\varphi=\pi$. The black dashed line corresponds to $\bar{\rho}_{\rm med}(n,t)$.} 
\label{Fig_Median}
\end{figure}

The apparent symmetry discussed in the previous section is also present in $\bar{\rho}_{\rm med}(n,t)$,
\begin{equation}
\bar{\rho}_{\rm med}(n,t)\equiv\frac{\bar{\rho}_{\sup}(n,t)+\bar{\rho}_{\inf}(n,t)}{2},
\end{equation} 
inherited from $\bar{\rho}_{\sup}(n,t)$ and $\bar{\rho}_{\inf}(n,t)$. The new function can be also retrieved from Eq.~\eqref{Main_Prob_Asymp} once one removes the sinusoidal term~\cite{NK03}:
\begin{eqnarray}
\bar{\rho}_{\rm med}(n,t)&=&\frac{1}{2 \pi t} \frac{1}{\nu(1-\nu)}\frac{\sin\theta}{\sqrt{\cos^2\theta-(2\nu-1)^2}}\nonumber \\
&\times&\Big[1-(2\nu-1) \left(\cos 2\eta +\sin 2\eta \tan\theta\cos\varphi \right)\Big].\nonumber \\
\label{Prob_Asymp_Med}
\end{eqnarray}
The appealing of $\bar{\rho}_{\rm med}(n,t)$ is twofold. On the one hand, it gives a smooth estimate of $\rho(n,t)$, see Fig.~\ref{Fig_Median}, much more concise than $\bar{\rho}(n,t)$. On the other hand, it serves as a starting point for the derivation of a stationary probability density function to which the exact PMF should converge as $t$ increases~\cite{GJS04}.

To this end, let us define the continuous variable $\varepsilon$, $\varepsilon\in(-1,+1)$,
\begin{equation}
\varepsilon\equiv \frac{2\nu-1}{\cos\theta},
\label{e_def}
\end{equation} 
which relates back to $n$ through the expression
\begin{equation*}
n=\frac{t}{2}\left(1+\varepsilon \cos\theta\right).
\end{equation*}
Therefore, for $t$ large enough we have
\begin{equation*}
\Delta n \sim \frac{t}{2}\cos\theta\, d\varepsilon,
\end{equation*}
and since we want
\begin{equation*}
\bar{\rho}_{\rm med}(n,t) \Delta n \sim \varrho(\varepsilon) d\varepsilon,
\end{equation*}
we obtain $\varrho(\varepsilon)$,
\begin{eqnarray}
\varrho(\varepsilon)&\equiv&\frac{\sin \theta}{\pi} \frac{1}{1- \varepsilon^2\cos^2\theta}\frac{1}{\sqrt{1-\varepsilon^2}}\nonumber \\
&\times&\Big[1-\varepsilon\left(\cos 2\eta \cos\theta +\sin 2\eta \sin\theta \cos\varphi\right)\Big].\nonumber \\
\label{st_dens}
\end{eqnarray}
The function thus defined is a probability density function because one has that $\sin\theta\geq 0$,
\begin{equation*}
\left|\cos 2\eta \cos\theta +\sin 2\eta \sin\theta \cos \varphi\right|\leq 1,
\end{equation*}
and
\begin{equation*}
\int_{-1}^{1}  \varrho(\varepsilon) d\varepsilon=1.
\end{equation*}


\section{Conclusion}
\label{Sec_Conclusion}

In this paper we have analyzed the evolution of a discrete-time quantum walk on a line when the coin operator shows the biggest generality from a physical point of view. In our approach the walker can either stand still in the place or proceed in a fixed direction but never move backward. This formalism is equivalent to the one used most, where the particle can travel in either direction, and therefore every formula or property can be easily translated from one setup to the other. 

Our main outcome is the derivation of closed-form expressions for the wave function that governs the probability of finding the particle at any given location, the probability mass function or PMF. 

These explicit results are the starting point for the analysis of some exact properties of the process, as its recursive nature or its achievable space symmetry, but also serve for the study of the asymptotic behavior of the quantum walker. 

In particular, we obtain two functions which estimate the upper and lower bounds of this PMF. These functions reveal the presence of an approximate symmetry in the process, a symmetry that is not shared by the exact solution. We also recover the general formula for the stationary probability density of the relative position of the walker, a law to which the process tends in the asymptotic limit. 

\begin{acknowledgments}
The author acknowledges partial support from the former Spanish Ministerio de Ciencia e Innovaci\'on under Contract No. FIS2009-09689, and from Generalitat de Catalunya, Contract No. 2009SGR417.
\end{acknowledgments}

\appendix
\section{Exact solution}
\label{App_Exact}
In this appendix we provide explicit details on the derivation of the closed-form expressions for the two components of the wave function given in the main text, Eqs.~\eqref{Sol_Psi0} and~\eqref{Sol_Psi1}, starting from the recursive formulas~\eqref{Rec_0} and~\eqref{Rec_1}. The approach that follows is similar to the one taken in our previous work~\cite{MM13}, and differs from the broadest method in the use of the discrete Fourier transform (DFT) instead of the discrete-time Fourier transform~\cite{VA12}. 

Let $f(n)$ be a given set of $N$ complex numbers, $n~\in~\{0,\ldots, N-1\}$, and denote its DFT by $\tilde{f}(r)$,
\begin{equation}
\tilde{f}(r)\equiv \sum_{n=0}^{N-1} f(n) e^{i 2 \pi r n/N},
\label{DFT}
\end{equation} 
for $r\in \{0,\ldots, N-1\}$. The $N$ complex quantities $\tilde{f}(r)$ thus defined keep exactly the same information contained in the original series, and therefore one can recover $f(n)$ from $\tilde{f}(r)$ by means of the so-called inverse DFT formula
\begin{equation}
f(n)\equiv \frac{1}{N}\sum_{r=0}^{N-1} \tilde{f}(r) e^{-i 2 \pi r n/N}.
\label{IDFT}
\end{equation}
The DFT is well suited to convert recursive relationships in the position domain, as Eqs.~\eqref{Rec_0} and~\eqref{Rec_1}, into a set of algebraic equations in the Fourier domain. 
In our case, however, the recursive formulas involve not only different locations but different instants of time, so we must carefully choose $f(n)$ and $N$ itself to preserve the overall coherence.

To this end, let us introduce the auxiliary time horizon $T$, $T\geq 0$, set $N\equiv T+1$, and consider the following definition for the DFT of $\psi_{0,1}(n,t)$, valid for any $t$, $t~\in~\{0,\ldots, T\}$,~\footnote{For notational convenience, $N$ and $T$ may alternate or even coexist in expressions in this appendix.}
\begin{equation}
\tilde{\psi}_{0,1}(r,t;T)\equiv \sum_{n=0}^{N-1} \psi_{0,1}(n,t) e^{i 2 \pi r n/N},
\label{DFT_Psi}
\end{equation}
where it is implicitly assumed that $\psi_{0,1}(n,t)=0$ for any $n\geq t+1$. This definition entails that $\tilde{\psi}_{0,1}(r,t;T)$ is an explicit function of $T$: that is, for a fixed value of $r$ and a fixed value of $t$, different choices of $T$ lead to different values for $\tilde{\psi}_{0,1}(r,t;T)$. Nonetheless, the final result that one gets after applying the corresponding inversion formula,
\begin{equation}
\psi_{0,1}(n,t)\equiv \frac{1}{N}\sum_{r=0}^{N-1} \tilde{\psi}_{0,1}(r,t;T) e^{-i 2 \pi r n/N},
\label{IDFT_Psi}
\end{equation}
does not depend on $T$ for a fixed choice of $n$ and $t$, as long as 
$0\leq n\leq t\leq N-1$.~\footnote{In fact, Eq.~\eqref{IDFT_Psi} is valid also in the range $n\in\{t+1,\ldots,N-1\}$. However, if one evaluates Eq.~\eqref{IDFT_Psi} for any of these values, one will obtain $\psi_{0,1}(n,t)=0$ identically.}
Therefore, the conclusion is that Eq.~\eqref{IDFT_Psi} will be valid for any $N$, $N\geq t+1$.

At this point we can safely move Eqs.~\eqref{Rec_0} and~\eqref{Rec_1} into the Fourier domain:
\begin{eqnarray}
\tilde{\psi}_{0}(r,t;T)&=&\cos\theta\, \tilde{\psi}_{0}(r,t-1;T)\nonumber \\
&+&e^{-i \varphi}\sin\theta\,\tilde{\psi}_1(r,t-1;T), \label{FRec_0}\\
\tilde{\psi}_{1}(r,t;T)&=&e^{i \varphi}\sin\theta\, e^{i 2 \pi r/N}\tilde{\psi}_{0}(r,t-1;T)\nonumber \\
&-&\cos\theta\, e^{i 2 \pi r/N}\tilde{\psi}_{1}(r,t-1;T).\label{FRec_1}
\end{eqnarray}
The initial values for $\tilde{\psi}_{0,1}(r,t;T)$ are $\tilde{\psi}_{0}(r,0;T)=\cos\eta$, $\tilde{\psi}_{1}(r,0;T)=\sin \eta $, for $r \in\{0,\ldots,N-1\}$. The resolution of Eqs.~\eqref{FRec_0} and~\eqref{FRec_1} can be tackled through standard matrix techniques, thus resulting in
\begin{widetext}
\begin{eqnarray}
\tilde{\psi}_{0}(r,t;T)&=&\frac{e^{-i\pi \frac{r}{N}}}{2\cos \orN}\left\{ (\lambda_+)^t \left[\left(\cos \theta-\lambda_-\right)\cos \eta + e^{-i \varphi}\sin\theta \sin \eta \right]
-(\lambda_-)^t \left[\left( \cos\theta -\lambda_+\right)\cos \eta + e^{-i \varphi}\sin\theta \sin \eta \right]\right\},\nonumber \\ \label{Sol_Fpsi0}
\end{eqnarray}
and
\begin{eqnarray}
\tilde{\psi}_{1}(r,t;T)&=&\frac{\left(\lambda_+ -\cos \theta\right)\left(\cos \theta-\lambda_-\right)e^{-i\pi \frac{r}{N}}}{2\cos \orN}\left\{ (\lambda_+)^t  \left[\frac{\cos \eta\, e^{i \varphi}}{\sin\theta} +\frac{\sin \eta}{\cos \theta-\lambda_-} \right]
-(\lambda_-)^t \left[\frac{\cos \eta\, e^{i \varphi}}{\sin\theta} -\frac{\sin \eta}{\lambda_+ -\cos \theta} \right]\right\},\nonumber \\ 
\label{Sol_Fpsi1}
\end{eqnarray}
\end{widetext}
with $\lambda_{\pm}$ functions of $r/N$,
\begin{eqnarray}
\lambda_+&\equiv&e^{-i(\orN -\pi \frac{r}{N})},\label{lambda_p}\\
\lambda_-&\equiv& -e^{i(\orN +\pi \frac{r}{N})},\label{lambda_m}
\end{eqnarray}
and where $\orN$ is an angle that, given $r$ and $N$, satisfies
\begin{equation}
\sin \orN =\cos\theta\,\sin  \frac{\pi r}{N}.
\label{App_omega_r}
\end{equation}
Note that, since $r\in\{0\ldots,N-1\}$, we have
\begin{equation*}
0\leq \sin \orN \leq \cos\theta\leq1, 
\end{equation*}
so, to prevent any uncertainty, we consider that $\orN$ is the only solution that Eq.~\eqref{App_omega_r} has in $[0,\frac{\pi}{2}]$. 

Now, we can simply introduce the value of $\tilde{\psi}_{0,1}(r,t;T)$ given in~\eqref{Sol_Fpsi0} and~\eqref{Sol_Fpsi1} into Eq.~\eqref{IDFT_Psi} and recover $\psi_{0,1}(n,t)$ after the computation of a finite sum. Note that, at this point, our procedure has yielded a closed-form expression for the wave function, thus we could simply stop here. However, we are going to simplify the final formulas as much as possible: in this way we can get a more detailed view of the properties of the solution.

To manage the complexity of this endeavor we analyze the particular case $\eta=0$ first. Let us begin with $\psi_{0}(n,t)$:
\begin{eqnarray}
\psi_{0}(n,t)&=&\frac{1}{N}\sum_{r=0}^{N-1}  
\frac{e^{-i\pi (2n +1) \frac{r}{N}}}{2\cos \orN}\left\{ (\lambda_+)^t\left(\cos \theta-\lambda_-\right)\right.\nonumber \\
&-&\left.(\lambda_-)^t \left(\cos \theta-\lambda_+\right)\right\}.
\end{eqnarray}
If we use Eq.~\eqref{App_omega_r} in conjunction with the definition of $\lambda_{\pm}$, Eqs.~\eqref{lambda_p} and~\eqref{lambda_m}, we can show that
\begin{equation}
e^{-i\pi r /N}\left(\cos\theta- \lambda_{\pm}\right)=\cos\theta \cos\frac{\pi r}{N} \mp \cos \orN
\label{cos_lambda}
\end{equation}
holds, and that consequently
\begin{eqnarray}
\psi_{0}(n,t)&=&\frac{1}{2N}\left[\left( \cos\theta+1\right) - (-1)^t\left( \cos\theta -1 \right) \right]\nonumber \\
&+&\frac{1}{2N}\sum_{r=1}^{N-1}  \frac{\cos \theta\cos\frac{\pi r}{N}}{\cos \orN} e^{-i\left[\pi (2n-t)\frac{r}{N}+ \orN t\right]}\nonumber \\
&+&\frac{1}{2N}\sum_{r=1}^{N-1}  e^{-i\left[\pi (2n-t)\frac{r}{N}+ \orN t\right]}\nonumber \\
&-&\frac{(-1)^t}{2N}\sum_{r=1}^{N-1}  \frac{\cos \theta\cos\frac{\pi r}{N}}{\cos \orN} e^{-i\left[\pi (2n-t)\frac{r}{N}- \orN t\right]}\nonumber \\
&+&\frac{(-1)^t}{2N}\sum_{r=1}^{N-1} e^{-i\left[\pi (2n-t)\frac{r}{N}- \orN t\right]}.
\end{eqnarray}
Here we have isolated in the first term all the contribution coming from the $r=0$ case. This step is necessary to rearrange the two last summations by introducing a new variable $s$, $s\equiv N-r$,
\begin{eqnarray}
\psi_{0}(n,t)&=&\frac{1}{2N}\left[\left( \cos\theta+1\right) - (-1)^t\left( \cos\theta -1 \right) \right]\nonumber \\
&+&\frac{1}{2N}\sum_{r=1}^{N-1}  \frac{\cos \theta\cos\frac{\pi r}{N}}{\cos \orN} e^{-i\left[\pi (2n-t)\frac{r}{N}+ \orN t\right]}\nonumber \\
&+&\frac{1}{2N}\sum_{r=1}^{N-1}  e^{-i\left[\pi (2n-t)\frac{r}{N}+ \orN t\right]}\nonumber \\
&+&\frac{1}{2N}\sum_{s=1}^{N-1}  \frac{\cos \theta\cos\frac{\pi s}{N}}{\cos \omega_\frac{s}{N}} e^{i\left[\pi (2n-t)\frac{s}{N}+ \omega_\frac{s}{N} t\right]}\nonumber \\
&+&\frac{1}{2N}\sum_{s=1}^{N-1} e^{i\left[\pi (2n-t)\frac{s}{N}+ \omega_\frac{s}{N} t\right]},
\end{eqnarray}
where the following identities
\begin{eqnarray*}
\orN&=&\omega_{\frac{s}{N}},\\ 
\cos\frac{\pi r}{N}&=&-\cos\frac{\pi s}{N},
\end{eqnarray*}
have been taken into account. Finally we find
\begin{eqnarray}
\psi_{0}(n,t)&=&\frac{1}{2N}\left[\left(\cos\theta+1\right) - (-1)^t\left(\cos\theta-1\right) \right]\nonumber \\
&+&\frac{1}{N}\sum_{r=1}^{N-1}  \left[1+\frac{\cos \theta \cos\frac{\pi r}{N}}{\cos \orN}\right] \nonumber \\ &\times& 
\cos\bigg[\pi (2n-t)\frac{r}{N}+ \orN t\bigg].
\end{eqnarray}
In the case of $\psi_{1}(n,t)$ we can proceed in a similar way. The result 
\begin{eqnarray}
\psi_{1}(n,t)&=&\frac{e^{i \varphi}\sin\theta }{N}\Bigg\{\frac{1 - (-1)^t}{2}\nonumber \\
&+&\sum_{r=1}^{N-1}  \frac{1 }{\cos \orN} \cos\bigg[\pi (2n-t-1)\frac{r}{N}+ \orN t\bigg]\Bigg\},\nonumber\\
\end{eqnarray}
is almost immediate once one realizes that
\begin{equation*}
\left(\lambda_{+}-\cos\theta\right)\left(\cos\theta- \lambda_{-}\right)e^{-i\pi r /N}=\sin^2\theta\,e^{i\pi r /N},
\end{equation*} 
an expression that combines Eqs.~\eqref{App_omega_r} and~\eqref{cos_lambda}.

Analogously, when $\eta=\frac{\pi}{2}$ we obtain
\begin{eqnarray}
\psi_{0}(n,t)&=&\frac{e^{-i\varphi}\sin\theta}{N}\Bigg\{\frac{1 - (-1)^t}{2}\nonumber \\
&+&\sum_{r=1}^{N-1}  \frac{1}{\cos \orN} \cos\bigg[\pi (2n-t+1)\frac{r}{N}+ \orN t\bigg]\Bigg\},\nonumber\\
\end{eqnarray}
and
\begin{eqnarray}
\psi_{1}(n,t)&=&\frac{1}{2N}\left[\left(1-\cos\theta\right) + (-1)^t\left(1+\cos\theta\right) \right]\nonumber \\
&+&\frac{1}{N}\sum_{r=1}^{N-1}  \left[1-\frac{\cos \theta \cos\frac{\pi r}{N}}{\cos \orN}\right] \nonumber \\ &\times& 
\cos\bigg[\pi (2n-t)\frac{r}{N}+ \orN t\bigg],
\end{eqnarray}
by simply applying the same ideas and intermediate formulas.

Finally, we can recover the general solution,  Eqs.~\eqref{Sol_Psi0} and~\eqref{Sol_Psi1}, through the superposition of these two cases.  

\section{Asymptotic expressions}
\label{App_Asymptotic}
In this appendix we obtain asymptotic expressions for $\psi_{0,1}(n,t)$ and $\rho(n,t)$, formulas with a restricted validity but which in turn are more compact and readable than the exact ones. 

We begin with a close analysis of the inner structure of the different pieces that compose Eqs.~\eqref{Sol_Psi0} and~\eqref{Sol_Psi1}.  The conclusion is that, in essence, we must find a way to approximate functions like $h(n,t)$, 
\begin{eqnarray}
h(n,t)&\equiv&\frac{\Xi(t)}{N}\nonumber\\&+& \frac{1}{N}\sum_{r=1}^{N-1} g(r/N) \cos\left[\Phi(n,r,t;T)+\epsilon \pi r/N\right],\nonumber\\
\end{eqnarray}
where 
\begin{equation}
\Phi(n,r,t;T)\equiv \pi (2n-t)r/N+ \orN t,
\end{equation}
and $\epsilon\in\{-1,0,1\}$. In every case $g(\cdot)$ is a smooth function, and therefore, the behavior of the cosine terms does determine the overall result of the sum. Due to the presence of $\orN$ within $\Phi(n,r,t;T)$, the argument of these cosine functions does not change linearly with $r$ but exhibits a maximum, and then the use an adapted version of the method of the stationary phase is the one most indicated in this case~\cite{MM13,CH53}: Only those terms for which $\Phi(n,r,t;T)$ attains its maximum are relevant, whereas the rest of them are negligible.   

To this end, let us first define $u\equiv r/N$, and $\nu\equiv n/t$, in terms of which we can rewrite $\Phi(n,r,t;T)$,
\begin{equation}
\Phi(\nu t,u (T-1) ,t;T)=\phi(\nu, u)t,
\end{equation}
with
\begin{equation}
\phi(\nu,u)\equiv \pi (2\nu-1) u+ \omega_u.
\end{equation}
Our next step is to consider the function $h(n,t)$ in the continuum limit, $N\to \infty$,
\begin{eqnarray}
h(n,t)&\sim&\int_0^{1}g(u) \cos\left[\phi(\nu,u) t+\epsilon \pi u\right] du\nonumber\\
&\sim& {\rm Re}\left\{\int_0^{1}g(u)e^{i \epsilon \pi u} e^{i\phi(\nu,u) t} du\right\},
\label{h_cont} 
\end{eqnarray}
and to expand $\phi(\nu,u)$ in the vicinity of $u_0$,
\begin{eqnarray*}
\phi(\nu,u)&\sim& \phi(\nu,u_0)+\frac{1}{2}\frac{\partial^2\phi(\nu,u_0)}{\partial u^2}(u-u_0)^2\\
&=&\phi_0(\nu)+\frac{1}{2}\phi''_0(\nu)(u-u_0)^2,
\end{eqnarray*}
where $u_0$ is the point for which, given $\nu$, $\phi(\nu,u)$ has its maximum: 
\begin{eqnarray}
\frac{\partial\phi(\nu,u_0)}{\partial u}&=& \pi (2\nu-1) +\frac{\pi \cos\theta\cos \pi u_0}{\sqrt{1-\cos^2\theta\sin^2 \pi u_0}}=0.\nonumber \\
\label{phi_sup}
\end{eqnarray}
From Eq.~\eqref{phi_sup} we have
\begin{equation}
\cos \pi u_0 =\frac{1-2\nu}{2\sqrt{\nu(1-\nu)}}\tan\theta,
\label{App_cos_pi_u_0}
\end{equation}
and
\begin{equation}
\sin \pi u_0 =\frac{1}{2\cos\theta}\sqrt{\frac{\cos^2\theta-(2\nu-1)^2}{\nu(1-\nu)}}.
\label{App_sin_pi_u_0}
\end{equation}
Equation~\eqref{App_sin_pi_u_0} shows us that the validity of the present approximation is restricted to values of $\nu$ for which one has $\cos^2\theta-(2\nu-1)^2\geq 0$; that is,
\begin{equation*}
\frac{1}{2}\left(1-\cos\theta\right)\leq\nu\leq\frac{1}{2}\left(1+\cos\theta\right),
\end{equation*}
since, by construction, $\cos\theta\geq0$. Also from Eqs.~\eqref{omega_r} and~\eqref{App_sin_pi_u_0} we get
\begin{eqnarray}
\sin \omega_0 &\equiv&\sin \omega_{u_0}=\frac{1}{2}\sqrt{\frac{\cos^2\theta-(2\nu-1)^2}{2\nu(1-\nu)}},\label{App_sin_omega_0} \\ 
\cos \omega_0 &\equiv&\cos \omega_{u_0}=\frac{\sin\theta}{2\sqrt{\nu(1-\nu)}}, \label{App_cos_omega_0}
\end{eqnarray}
expressions that will be helpful in forthcoming derivations.
 
Now we can fully evaluate Eq.~\eqref{h_cont} under the above premises:
\begin{eqnarray}
h(n,t)&\sim& {\rm Re}\left\{\int_0^{1}g(u)e^{i \epsilon \pi u} e^{i\phi(\nu,u) t} du\right\} \nonumber\\
&\sim& {\rm Re}\left\{\int_0^{1}g(u_0)e^{i \epsilon \pi u_0} e^{i t\left[\phi_0(\nu)+\frac{1}{2}\phi''_0(\nu)(u-u_0)^2\right]} du\right\}\nonumber \\
&\sim& {\rm Re}\left\{g(u_0)e^{i \left[\epsilon \pi u_0+\phi_0(\nu) t\right]}\int_{-\infty}^{\infty} e^{\frac{i t}{2}\phi''_0(\nu)(u-u_0)^2} du\right\}\nonumber \\
&=&\sqrt{\frac{2 \pi}{t|\phi''_0(\nu)|}} g(u_0) \cos\left[\phi_0(\nu) t-\frac{\pi}{4}+\epsilon \pi u_0\right], \nonumber\\
\label{h_result}
\end{eqnarray}
with
\begin{eqnarray}
\phi''_0(\nu) &=&-4 \pi^2 \nu(1-\nu)\frac{\sqrt{\cos^2\theta-(2\nu-1)^2}}{\sin\theta}.
\end{eqnarray}
The approximate versions of Eqs.~\eqref{Sol_Psi0} and~\eqref{Sol_Psi1} are
\begin{eqnarray}
\psi_{0}(n,t)&\sim&\frac{\cos\eta}{\sqrt{t}} \sqrt{\frac{2(1-\nu)\sin\theta}{\pi \nu\sqrt{\cos^2\theta-(2\nu-1)^2}}} \nonumber\\
&\times&\cos\left[\phi_0(\nu) t-\frac{\pi}{4}\right]\nonumber \\
&+&\frac{e^{-i \varphi}\sin\eta}{\sqrt{t}} \sqrt{\frac{2\sin\theta}{\pi \sqrt{\cos^2\theta-(2\nu-1)^2}}} \nonumber \\
&\times&\cos\left[\phi_0(\nu) t-\frac{\pi}{4}+\pi u_0\right], \label{Sol_App0}
\end{eqnarray}
and
\begin{eqnarray}
\psi_{1}(n,t)&\sim&\frac{e^{i\varphi}\cos\eta }{\sqrt{t}} \sqrt{\frac{2\sin\theta}{\pi \sqrt{\cos^2\theta-(2\nu-1)^2}}} \nonumber \\
&\times&\cos\left[\phi_0(\nu) t-\frac{\pi}{4}-\pi u_0\right]\nonumber \\
&+& \frac{\sin\eta}{\sqrt{t}} \sqrt{\frac{2\nu\sin\theta}{\pi (1-\nu)\sqrt{\cos^2\theta-(2\nu-1)^2}}}\nonumber \\
&\times& \cos\left[\phi_0(\nu) t-\frac{\pi}{4}\right],
\label{Sol_App}
\end{eqnarray}
and they follow from Eq.~\eqref{h_result} once one realizes, cf. Eqs.~\eqref{App_cos_pi_u_0} and~\eqref{App_cos_omega_0}, that if 
\begin{equation*}
g(u)= 1+\frac{\cos \theta \cos \pi u}{\cos \omega_u},
\end{equation*}
one gets
\begin{equation*}
g(u_0)= 2(1-\nu);
\end{equation*}
if
\begin{equation*}
g(u)=  \frac{\sin\theta}{\cos \omega_u},
\end{equation*}
one has
\begin{equation*}
g(u_0)= 2\sqrt{\nu(1-\nu)};
\end{equation*}
and finally if
\begin{equation*}
g(u)= 1-\frac{\cos \theta \cos \pi u}{\cos \omega_u},
\end{equation*}
one obtains
\begin{equation*}
g(u_0)= 2\nu.
\end{equation*}

To derive the asymptotic expression for $\rho(n,t)$ one has to calculate $\left|\psi_{0,1}(n,t)\right|^2$ first:
\begin{widetext}
\begin{eqnarray}
\left|\psi_0(n,t)\right|^2&\sim&\frac{1}{t} \frac{2\sin\theta}{\pi \nu\sqrt{\cos^2\theta-(2\nu-1)^2}} 
\Bigg\{(1-\nu)\cos^2\eta\cos^2\left[\phi_0(\nu) t-\frac{\pi}{4}\right]+\nu\sin^2\eta\cos^2\left[\phi_0(\nu) t-\frac{\pi}{4}+ \pi u_0\right]\nonumber \\
&+&\sqrt{\nu(1-\nu)}\sin 2\eta\cos \varphi \cos\left[\phi_0(\nu) t-\frac{\pi}{4}\right]\cos\left[\phi_0(\nu) t-\frac{\pi}{4}+ \pi u_0\right]\Bigg\},\label{Psi_0_App}\\
\left|\psi_1(n,t)\right|^2&\sim&\frac{1}{t} \frac{2\sin\theta}{\pi (1-\nu)\sqrt{\cos^2\theta-(2\nu-1)^2}} 
\Bigg\{(1-\nu)\cos^2\eta\cos^2\left[\phi_0(\nu) t-\frac{\pi}{4}-\pi u_0\right]+\nu\sin^2\eta\cos^2\left[\phi_0(\nu) t-\frac{\pi}{4}\right]\nonumber \\
&+&\sqrt{\nu(1-\nu)}\sin 2\eta\cos \varphi \cos\left[\phi_0(\nu) t-\frac{\pi}{4}\right]\cos\left[\phi_0(\nu) t-\frac{\pi}{4}-\pi u_0\right]\Bigg\},
\label{Psi_1_App}
\end{eqnarray}
then use the following trigonometric identities,
\begin{eqnarray*}
&2&\cos^2\left[\phi_0(\nu) t-\frac{\pi}{4}\right]= 1+ \sin\left[2\phi_0(\nu) t\right],\\
&2&\cos\left[\phi_0(\nu) t-\frac{\pi}{4}\right]\cos\left[\phi_0(\nu) t-\frac{\pi}{4}\pm \pi u_0\right]=\cos \pi u_0 +\cos \pi u_0 \sin\left[2\phi_0(\nu) t\right]\pm \sin \pi u_0 \cos\left[2\phi_0(\nu) t\right],\\
&2&\cos^2\left[\phi_0(\nu) t-\frac{\pi}{4}\pm \pi u_0\right]=1+\cos 2\pi u_0 \sin\left[2\phi_0(\nu) t\right]\pm \sin 2\pi u_0 \cos\left[2\phi_0(\nu) t\right],
\end{eqnarray*}
in coordination with Eqs.~\eqref{App_cos_pi_u_0} to~\eqref{App_cos_omega_0}, to ultimately obtain
\begin{eqnarray}
\rho(n,t)&\sim&\frac{1}{2 \pi t} \frac{1}{\nu(1-\nu)}\frac{\sin\theta}{\sqrt{\cos^2\theta-(2\nu-1)^2}} \Bigg\{1-(2\nu-1) \left(\cos 2\eta +\sin 2\eta \tan\theta \cos \varphi\right)\nonumber\\
&+&(2\nu-1) \left[\frac{2\nu-1}{\cos^2\theta}-\cos 2\eta -\sin 2\eta \tan\theta \cos \varphi\right]\sin \left[2\phi_0(\nu) t\right]\nonumber\\
&+&(2\nu-1) \sqrt{1-\frac{(2\nu-1)^2}{\cos^2\theta}}\bigg[\cos 2\eta \tan\theta-\sin 2\eta \cos \varphi\bigg]\cos\left[2\phi_0(\nu) t\right]\Bigg\}\equiv\bar{\rho}(n,t).
\label{Prob_Asymp_App}
\end{eqnarray}
\end{widetext}
This expression leads to Eq.~\eqref{Main_Prob_Asymp} after a straightforward trigonometric transformation.

\end{document}